\documentclass{aa}
\usepackage{graphicx}
\usepackage{longtable}
\begin{document}
\title{The Initial Helium Abundance of the Galactic Globular Cluster
System 
\thanks{Based on observations with the NASA/ESA {\it Hubble Space
Telescope}, obtained at the Space Telescope Science Institute, which is operated by
AURA, Inc., under NASA contract NAS5-26555, and on observations retrieved with the
ESO ST-ECF Archive.}
}
\author{
M. Salaris\inst{1,2} 
\and
M. Riello\inst{3,4}
\and
S. Cassisi\inst{5}
\and
G. Piotto\inst{3}
}

\offprints{Maurizio Salaris}

\institute{
Astrophysics Research Institute, Liverpool John Moores University,
Twelve Quays House, Egerton Wharf, Birkenhead, CH41 1LD, UK
\email{ms@astro.livjm.ac.uk}
\and
Max-Planck-Institut f\"ur Astrophysik, Karl-Schwarzschild-Strasse 1,
85758, Garching, Germany
\and
Dipartimento di Astronomia, Vicolo dell'Osservatorio, 2, I-35122 Padova, Italy
\email{piotto,riello@pd.astro.it}
\and
INAF - Osservatorio Astronomico di Padova, vicolo dell'Osservatorio 5,
I--35122 Padova, Italy
\and 
INAF - Osservatorio Astronomico di Collurania, Via M. Maggini, Teramo,
I--64100, Italy
\email{cassisi@te.astro.it}
}

\date{Received ; accepted}

\abstract{
In this paper we estimate the initial He content in about 30\% of the
Galactic 
globular clusters (GGCs) from new star counts we have performed on the recently
published HST snapshot database of Colour Magnitude Diagrams (Piotto et
al. 2002). 
More specifically, we use the so-called $R$-parameter and estimate the 
He content from a theoretical calibration based on a recently updated
set of stellar evolution models.
We performed an accurate statistical analysis in order to assess whether
GGCs show a statistically significant spread in 
their initial He abundances, and whether there is a
correlation with the cluster metallicity.
As in previous works on the subject, 
we do not find any significant dependence of the He abundance on the
cluster metallicity; this provides an important constraint 
for models of Galaxy formation and evolution.
Apart from GGCs with the bluest Horizontal Branch morphology, 
the observed spread in the individual helium abundances is 
statistically compatible with
the individual errors. This means that either there is no
intrinsic abundance spread among the GGCs, or that this is 
masked by the errors. In the latter case we have estimated  
a firm 1$\sigma$ upper limit of 0.019 to the possible intrinsic spread.
In case of the GGCs with the
bluest Horizontal Branch morphology we detect a significant spread 
towards higher abundances inconsistent with the individual errors; this
can be fully explained by additional
effects not accounted for in our theoretical calibrations, which do not
affect the abundances estimated for the clusters with redder
Horizontal Branch morphology.
In the hypothesis that the intrinsic dispersion on the individual He
abundances is zero, taking into account the errors on 
the individual $R$-parameter estimates, as well as the uncertainties
on the cluster metallicity scale and theoretical
calibration, we have determined an initial He abundance mass fraction
$Y_{\rm GGC}=0.250\pm0.006$. This value is in perfect agreement with
current estimates based on Cosmic Microwave Background radiation 
analyses and cosmological nucleosynthesis
computations.
}
\authorrunning{M. Salaris et al.}
\titlerunning{Initial He abundance of globular clusters}
   \maketitle

\section{Introduction}

The determination of the initial He abundance of Galactic Globular
Cluster (GGC) stars plays an important role in astrophysics, 
because of its wide-ranging implications.

First of all, due to the fact that GGC stars are the oldest objects
in the Galaxy, their initial He abundance ($Y_{\rm GGC}$, where $Y$
denotes the mass fraction of He) should mirror 
the primordial He abundance ($Y_p$) produced 
during the big bang nucleosynthesis.
Secondly, the presence (or absence) of gradients in $Y_{\rm GGC}$ as a
function of the cluster metallicity provides information about Galactic
nucleosynthesis and evolution. Thirdly, the existence of an 
abundance spread at constant metallicity would strenghten the case
for He being the so-called 'second parameter' (beside metallicity) that determines
the morphology of the Horizontal Branch (HB) in the GGC
Colour-Magnitude-Diagrams (CMDs -- Sandage \& Wildey~1967).

Empirical estimates of $Y_{\rm GGC}$, are necessarily indirect, since     
He lines are not detectable in GGC star spectra,     
apart from the case of hot HB objects, whose     
atmospheres are however affected by gravitational settling and     
radiative levitation, which strongly alter their initial chemical      
stratification (see, e.g., Michaud et al.~1983).      
$Y_{\rm GGC}$ estimates take advantage of results from stellar     
evolution theory, which show that 
the evolution of low mass Population~II stars is     
affected by the initial He content. The so-called     
$R$-parameter (Iben~1968; Buzzoni et al.~1983; Caputo et al.~1987), 
defined as the number ratio of HB      
to Red Giant Branch (RGB) stars brighter than the HB level      
($R=N_{HB}/N_{RGB}$), is employed to determine $Y_{\rm GGC}$. At a given     
metallicity, a higher initial He-content implies a brighter HB and,     
in turn, a lower value of $N_{RGB}$ ($N_{HB}$ is only slightly affected),      
with the consequent increase of $R$.     
Other parameters derived from stellar evolution can      
also be employed (see, e.g., the discussions     
in Sandquist~2000; Zoccali et al.~2000), but they are better suited to     
determine relative He abundances than absolute ones.   

In Cassisi, Salaris \& Irwin~(2003, hereafter Paper~I)
we have shown how $R$-parameter determinations in two
samples of GGCs (Sandquist~2000, 43 objects; Zoccali et al.~2000, 
26 objects), coupled with a theoretical calibration obtained from updated
stellar evolution models, provide a constant value of 
$Y_{\rm GGC}$, with no trend with respect to [Fe/H]. The weighted average 
of the individual cluster abundances provides $Y_{\rm GGC}$=0.244$\pm$0.006 or 
$Y_{\rm GGC}$=0.243$\pm$0.006, the negligible difference being due to the
choice of the observational sample. This value of $Y_{\rm GGC}$
is in agreement -- within the errors -- with
recent determinations of the cosmological baryonic     
matter density ($\Omega_b$) from the Cosmic Microwave 
Background (CMB) power spectrum      
obtained by the BOOMERANG, DASI, MAXIMA and WMAP experiments     
(e.g. Pryke et al.~2002; \"Odman et al.~2003; 
Sievers et al.~2003; Spergel et al.~2003) which, coupled to 
BBN calculations (Burles, Nollett \& Turner~2001), do consistently 
provide $Y_p$=0.248$\pm$0.001. It is  important to 
recall that earlier studies (Sandquist~2000; Zoccali et al.~2000) that employed
older generations of stellar models, found $Y_{\rm GGC}$$\sim$0.20, in complete       
disagreement with the CMB constraint and spectroscopic determinations
from HII regions ($Y_p$=0.234$\pm$0.002 according to  
Olive, Steigman \& Skillman~1997; $Y_p$=0.244$\pm$0.002 according
to Izotov \& Thuan~1998).

As for the abundance spread at a given [Fe/H], the results of Paper~I were
contradictory, in the sense that $R$-parameter values from the 
Zoccali et al. (2000) sample show a spread that is completely
consistent with the measurement errors, whereas the sample by      
Sandquist~(2000) shows an 
intrinsic additional 1$\sigma$ spread in the He abundance of $\sim$0.03.

In this paper we measured new $R$-parameter values for a sample of 57 GGCs
from the database of HST GGC observations
by Piotto et al.~(2002). We use the HST flight photometric
system (without reddening corrections) instead of the Johnson photometric bands. 
This approach avoids potential subtle errors in the estimate of the
visual magnitudes of the HB. In fact, the transformation to the
Johnson system requires the knowledge of the cluster reddening,
therefore the accuracy of the observed Johnson magnitudes depends on
the accuracy of the cluster reddening estimate. 

Our determinations of $R$-parameter values constitute 
the largest existing homogeneous
database, encompassing about 30\% of
the total GGC population. It is therefore extraordinarily well suited
to address in more detail the question
of the intrinsic spread in the individual GGC He abundances, and  
the ratio of He enrichment with metallicity enrichment
($\Delta Y/\Delta Z$). With regard to the
latter quantity, its evaluation is extremely important 
because of the connections with stellar yields and mechanism of
formation of the GGCs, and the determination of their absolute ages.
In addition, it is relevant to make a comparison  
with the value $\Delta Y/\Delta Z$ between
$\approx$ 6 and 12 (depending on the sample adopted, and with the
assumption that the Oxygen abundance is a good tracer of total metallicity)  
estimated by Olive et al.~(1997) for extragalactic HII regions, and
the value $\sim$ 3 estimated by Izotov \& Thuan~(2003).  
We will also address the hypothesis put forward by D'Antona et
al.~(2002), which suggests the existence of an He enriched stellar
component within individual clusters, due to chemical 
pollution by the ejecta of
massive asymptotic giant branch stars; in particular, we will assess
if this phenomenon might seriously affect our determination of 
$Y_{\rm GGC}$. 

In Sect.~2 we briefly describe the cluster sample and the theoretical
models employed. Section~3 presents our results for the
$R$-parameter measurements and the determination of $Y_{\rm GGC}$. 
A summary and conclusions follow in Sect.~4.

\section{Cluster sample and theoretical stellar models}

We have employed our large photometric database of 74 GGCs
observed in the HST $B$ (F439W) and $V$ (F555W) bands with the WFPC2
(Piotto et al.~2002). The observations, pre-processing, photometric
reduction and calibration of the instrumental magnitudes to the HST
flight system, as well as the artificial star experiments performed to
derive the star count completeness are described in Piotto et al.~(2002).
For each cluster we measured a number of stars that ranges from a few
thousand to $\approx$ 47000 (in case of NGC~6388). 
For 57 of our clusters we have been able to measure $R$.
The empirical definition of $R$ is the same as in Zoccali et
al.~(2000), i.e., the number of RGB stars is computed starting from the
level of the observed F555W magnitude of the Zero Age HB (ZAHB; the
lower envelope of the observed HB star distribution). The
determination of the ZAHB level has been discussed in Riello et
al.~(2003) and Recio-Blanco et al.~(2004, in preparation).
The error on the observed $R$ values is computed by combining in
quadrature the Poisson error associated with the measurement of 
$N_{HB}$ and $N_{RGB}$, and the contribution due to the uncertainty on
the ZAHB level (which translates into an additional uncertainty on 
$N_{RGB}$, hence on $R$). Usually, this second contribution is much
smaller than the Poisson error associated with the number counts 
$N_{HB}$ and $N_{RGB}$.
As for the GGC [Fe/H] values, we take into account current
uncertainties on this quantity by employing both
the Carretta \& Gratton~(1997 -- CG97) and Zinn \& West~(1984 -- ZW84)
scales (we assumed an error of $\pm$0.15 dex on the individual abundances).
The observational data relevant to our analysis are summarized in Table~\ref{Rdata}.

The theoretical models and isochrones needed to calibrate the relationship
between $R$ and $Y$ are the same as in Paper~I, computed for a range
of [Fe/H] and $Y$ values.
A full description of these models will be presented in a
forthcoming paper (Pietrinferni et al.~2004, submitted to ApJ) however,
the basic physical ingredients relevant for the computation of 
the theoretical $R$ values have already been discussed in Paper~I. 
Here, we briefly recall that our models have been computed with 
the $\alpha$-enhanced metal distribution given in Salaris \& Weiss~(1998), with
$<$[$\alpha$/Fe]$>$=0.4. The radiative opacities 
(Alexander \& Ferguson~1994; Iglesias \& Rogers~1996)  
have been computed specifically for
our adopted metal mixture, and the electron conduction opacities 
are from Potekhin~(1999). The nuclear reaction rates from the NACRE
database (Angulo et al.~1999) have been employed, with the exception
of the $^{12}$C$(\alpha,\gamma)^{16}$O reaction. For this reaction we employ the      
more accurate recent determination by Kunz et al.~(2002), based on $\gamma$     
angular distribution measurements of $^{12}$C$(\alpha,\gamma)^{16}$O     
and a consistent ${\mathbf R}$-matrix analysis of the process. The      
claimed relative uncertainty of this new rate is half of the     
uncertainty quoted in previous determinations.   
The relevant energy loss rates from plasma-neutrino     
processes have been taken from Haft et al.~(1994).
We have used the equation of state by Irwin et al.~(2004, in preparation)\footnote{
the equation of state code is made publicly available at     
ftp://astroftp.phys.uvic.ca/pub/irwin/eos/code/ eos\_demo\_fortran.tar.gz} 
computed for our actual $\alpha$-enhanced metal mixture.

Table~\ref{compmodels} compares, as an example, the adopted ZAHB
level and corresponding He core mass, HB and RGB evolutionary times
entering our present theoretical calibration of the $R$-parameter, with
the calibration by Zoccali et al.~(2000), for $Y$=0.245 and two
selected metallicities. The data for the Zoccali et al.~(2000) model
have been obtained by interpolating within their computed grid, since
they did not specifically compute models for $Y$=0.245. By simply
comparing the evolutionary timescales that enter the $R$ calibration one
notices a reduction of $\sim$20\% of the value of $R$ when passing
from Zoccali et al.~(2000) models to our new ones, as discussed in Paper~I. 
As a further comparison to highlight the change in the
theoretical calibration of $R$ due to
the improvements in stellar models, we also
display in the same table analogous data from Cassisi et al.~(1998)
computations, for $Y$=0.245. These models were a sort of intermediate 
step between the models used in Zoccali et al.~(2000) and in Paper~I, 
and have never been used to determine $Y_{GGC}$. A calibration based on
Cassisi et al.~(1998) models reduces the value of $R$ 
by $\sim$10\% at Z=0.001 with respect to Zoccali et 
al.~(2000), whereas at Z=0.006 the reduction is by only about 2\%.

The bolometric luminosities and effective temperatures of the models
have been converted to the HST flight system using the
transformations by Origlia \& Leitherer~(2000), based on the model
atmospheres by Bessel et al.~(1998). 
In Paper~I we have also shown how our treatment of the semiconvective regions  
during the He burning phase does agree with independent observational
constraints coming from the 
$R_2$ parameter measured in a sample of GGCs by Sandquist~(2000 -- see Paper~I for details).

The theoretical values of $R$     
are computed by considering the RGB evolutionary time of the RGB mass
populating the relevant isochrone, and the HB lifetime of a     
star populating the middle of the RR~Lyrae instability strip     
(log$(T_{eff})$=3.85). This is strictly adequate only for those clusters     
with an HB populated at the RR~Lyrae instability strip and redward     
(increasing total stellar mass), since the HB evolutionary timescales are     
basically unchanged when moving from the instability strip towards the red     
(see Paper~I and references therein). However, stars populating the
bluer part of the HB do show different evolutionary times, which increase     
for decreasing total stellar mass (bluer colour). At the bluest end of
a typical HB the increase of the HB evolutionary time with respect to
the RR~Lyrae strip counterpart can amount 
to about 20 \% (see, e.g., Zoccali et al.~2000).


\begin{table} 
\caption[]{Comparison of the ZAHB levels, He core masses (in solar
mass units), HB evolutionary times (in Myr) and RGB evolutionary times
at brighter magnitudes than the ZAHB (in Myr) between the models used
in Zoccali et al.~(2000), in Paper~I, and the models by Cassisi et
al.~(1998), for two selected metallicities
(see text for details).}
\begin{tabular}{lccccc} \hline 
            Zoccali et al.~(2000)
            & $M_V(ZAHB)$ 
            & $M_c(He)$
            & $t_{HB}$
            & $t_{RGB}$\\
\hline
Z=0.001  & 0.54   & 0.499  & 102 & 63\\
Z=0.006  & 0.72   & 0.493  & 112 & 55\\
\hline
Cassisi et al.~(1998) & & & & \\
\hline
Z=0.001 & 0.49  & 0.500   & 85   & 58 \\
Z=0.006 & 0.68  & 0.493   & 94   & 47 \\
\hline
Paper~I & & & & \\
\hline
Z=0.001  & 0.58   & 0.495  &  88  & 65\\
Z=0.006  & 0.74   & 0.489  &  93  & 56\\
\hline 				
\label{compmodels} 
\end{tabular} 
\end{table} 

\subsection{The effect of He enriched subpopulations in individual clusters}

Another effect that might bias our determination of $Y_{\rm GGC}$ is 
the He enrichment in some GGC proposed by D'Antona et al.~(2000). More in
detail, to explain the observed CNO abundance anomalies and the
extended blue HB tails in some GGCs, D'Antona et al.~(2002) suggested
the existence of an He enriched stellar component within the
clusters. This enrichment is due to chemical pollution by the ejecta
of massive asymptotic giant branch stars belonging to the cluster;
this chemical pollution may explain the mentioned CNO
anomalies and HB colours. In Riello et al.~(2003) we showed 
that this supposed enrichment does not alter the brightness of the 
RGB bump in the luminosity function of GGCs. Here we estimate the effect on the He
abundance estimated through the $R$ parameter. As in Riello et
al.~(2003), and similarly to D'Antona et al.~(2002), we considered a
cluster stellar population composed of 64\% of stars with a given 
initial He abundance as adopted in our models -- here denoted as
'normal' -- , and 36\% of stars with
an initial $Y$ randomly distributed between the 'normal' value 
and a mass fraction 0.06 larger. In general, for any morphology of the
HB, the level of the HB that enters the definition of $R$ is still determined 
by the 64\% of objects with 'normal' abundance; 
$N_{HB}$ for the whole population tends to increase, whereas
$N_{RGB}$ decreases, 
increasing the helium content estimate. 
When using our present standard calibration of $R$ as a
function of $Y$ in He-enhanced clusters, we would overestimate
the initial He mass fraction by at most 0.005--0.006 
for clusters where the HB instability strip and the region 
redwards of this strip are populated, whereas for blue clusters the effect is
an overestimate of at most 0.01--0.02. Most probably the degree of He
enrichment (if it is real) will vary from cluster to cluster (for the
clusters affected by this phenomenon), and the real overestimate will
be between zero and the values quoted above.


\begin{table*} 
\renewcommand{\arraystretch}{0.7} 
\caption[]{Cluster data. The columns display, respectively, 
cluster name, measured value of $R$ with associated error, [Fe/H] on
the ZW84 and CG97 scale, and value of $HB_{type}$ (see text for details)}
\begin{tabular}{lccccr} \hline 
            Cluster
            & $R$ 
            & $\sigma(R)$
            & $\rm [Fe/H]_{ZW84}$
            & $\rm [Fe/H]_{CG97}$
            & $HB_{type}$\\
\hline
     IC4499    &  1.351  &    0.302  &    $-$1.50  &    $-$1.27  &     0.11\\
     NGC~104   &  1.607   &   0.157 &     $-$0.71  &    $-$0.70  &    $-$0.99\\
     NGC~362   &  1.358   &   0.193  &    $-$1.27  &    $-$1.15  &    $-$0.87\\
     NGC~1261  &  1.208   &   0.170  &    $-$1.31  &    $-$1.10  &    $-$0.71\\
     NGC~1851  &   1.457  &    0.163  &    $-$1.36 &     $-$1.14  &    $-$0.36\\
     NGC~1904  &   2.055   &   0.251  &    $-$1.69  &    $-$1.37  &     0.89\\
     NGC~2808  &   1.598  &    0.139  &    $-$1.37  &    $-$1.15  &    $-$0.49\\
     NGC~3201  &   1.136  &    0.319  &    $-$1.37  &    $-$1.15  &     0.08\\
     NGC~4147  &   1.767  &    0.408  &    $-$1.80 &     $-$1.59  &     0.55\\
     NGC~4372  &   1.111  &    0.366  &    $-$2.08  &    $-$1.94 &      1.00\\
     NGC~4590  &   0.854   &   0.200  &    $-$2.09 &     $-$1.99  &     0.17\\
     NGC~4833  &   2.189  &    0.405  &    $-$1.86 &     $-$1.58  &     0.93\\
     NGC~5024  &   1.477  &    0.159  &    $-$2.04 &     $-$1.89   &    0.81\\
     NGC~5634  &  1.433  &    0.184  &    $-$1.82  &    $-$1.61  &     0.91\\
     NGC~5694  &  1.537   &   0.167  &    $-$1.92  &    $-$1.73  &     1.00\\
     NGC~5824  &  1.415   &   0.092  &    $-$1.87  &    $-$1.67  &     0.79\\
     NGC~5904  &    1.199  &    0.146  &    $-$1.40  &    $-$1.11  &     0.31\\
     NGC~5927  &   1.561  &    0.199  &    $-$0.30  &    $-$0.62  &    $-$1.00\\
     NGC~5946  &   1.321  &    0.171 &     $-$1.37  &    $-$1.15  &     0.71\\
     NGC~5986  &   1.423  &    0.144  &    $-$1.67  &    $-$1.44  &     0.97\\
     NGC~6093  &   1.031  &    0.144  &    $-$1.67  &    $-$1.44  &     0.97\\
     NGC~6139  &   1.244  &    0.121  &    $-$1.65  &    $-$1.42  &     0.91\\
     NGC~6171  &   1.560  &    0.432  &    $-$0.99 &     $-$0.87  &    $-$0.73\\
     NGC~6205  &   1.719  &    0.197  &    $-$1.65 &     $-$1.39  &     0.97\\
     NGC~6218  &   1.366  &    0.292  &    $-$1.61  &    $-$1.37  &     0.97\\
     NGC~6229  &   1.485  &    0.140  &    $-$1.54  &    $-$1.30  &     0.24\\
     NGC~6235  &   0.949  &    0.218  &    $-$1.40 &     $-$1.17  &     0.89\\
     NGC~6266  &    1.662  &    0.153  &    $-$1.28 &     $-$1.07  &     0.32\\
     NGC~6273  &    1.554  &    0.136  &    $-$1.68 &     $-$1.45  &     0.97\\
     NGC~6284  &   1.210  &    0.157  &    $-$1.40  &    $-$1.17  &     0.83\\
     NGC~6287  &    1.519 &     0.273  &    $-$2.05  &    $-$1.90  &     0.98\\
     NGC~6293  &    1.351 &     0.190  &    $-$1.92  &    $-$1.73  &     0.90\\
     NGC~6304  &    1.818 &     0.306  &    $-$0.59  &    $-$0.68  &    $-$1.00\\
     NGC~6342  &    1.771 &     0.375  &    $-$0.62  &    $-$0.69  &    $-$1.00\\
     NGC~6356   &   1.658 &     0.153  &    $-$0.62  &    $-$0.69  &    $-$1.00\\
     NGC~6362   &   1.429 &     0.367  &    $-$1.08  &    $-$0.96  &    $-$0.58\\
     NGC~6388   &   2.130 &     0.121 &     $-$0.74  &    $-$0.74  &    $-$0.67\\
     NGC~6441   &   1.854 &     0.113  &    $-$0.59   &   $-$0.68 &     $-$0.78\\
     NGC~6522   &   1.183  &    0.159  &    $-$1.44  &    $-$1.21  &     0.71\\
     NGC~6544   &   1.500 &     0.395 &     $-$1.56  &    $-$1.32  &     1.00\\
     NGC~6569   &   1.547 &     0.202  &    $-$0.86  &    $-$0.80  &    $-$0.76\\
     NGC~6584   &   1.217  &    0.235  &    $-$1.54  &    $-$1.30  &   $-$0.15\\
     NGC~6624   &   1.605 &     0.243  &    $-$0.35  &    $-$0.63  &   $-$1.00\\
     NGC~6637    &  2.060 &     1.172 &     $-$0.59  &    $-$0.69  &   $-$1.00\\
     NGC~6638    &  1.228  &    0.186 &     $-$1.15  &    $-$0.97  &   $-$0.30\\
     NGC~6642   &   1.319  &    0.265  &    $-$1.29  &    $-$1.08  &    0.29\\
     NGC~6652   &   1.512  &    0.309  &    $-$0.89   &   $-$0.81  &   $-$1.00\\
     NGC~6681  &    1.755  &    0.283  &    $-$1.51  &    $-$1.27   &   0.96\\
     NGC~6717  &    0.722  &    0.267  &    $-$1.32  &    $-$1.10  &    0.98\\
     NGC~6723   &   2.383  &    1.367  &    $-$1.09  &    $-$0.93  &   $-$0.08\\
     NGC~6760   &   1.244  &    0.145  &    $-$0.52  &    $-$0.66  &   $-$1.00\\
     NGC~6864   &   1.712  &    0.194  &    $-$1.32  &    $-$1.10   &  $-$0.07\\
     NGC~6934   &   1.621  &    0.218  &    $-$1.54  &      $-$1.30  &    0.25\\
     NGC~6981   &   1.088  &    0.205  &    $-$1.54  &    $-$1.30  &    0.14\\
     NGC~7078   &   1.883  &    0.175  &    $-$2.15  &    $-$2.12   &   0.67\\
     NGC~7089   &   1.455  &    0.183  &    $-$1.62  &    $-$1.39   &   0.96\\
     NGC~7099   &   2.667  &    0.531  &    $-$2.13  &    $-$1.91   &   0.89\\
\hline 				
\label{Rdata} 
\end{tabular} 
\end{table*} 

\begin{figure}
\resizebox{\hsize}{!}{\includegraphics{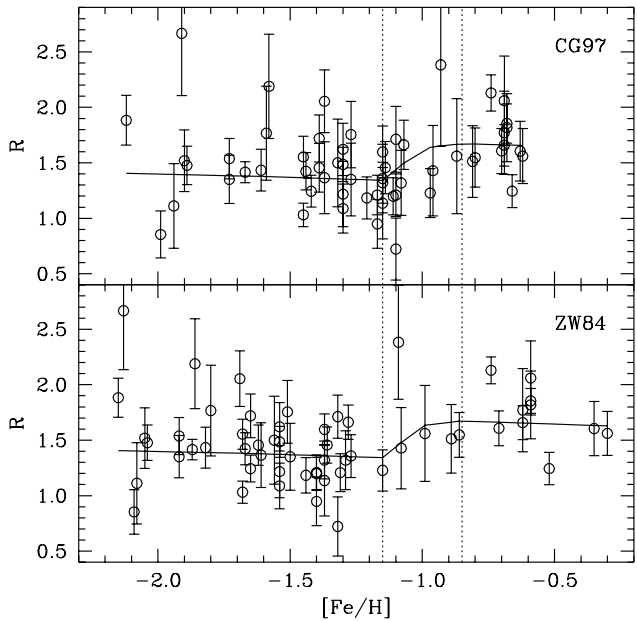}}
\caption{Observed $R$-parameter values as a function of the cluster
[Fe/H] for the two adopted metallicity scales. The theoretical calibration for $Y$=0.245
and a cluster age of 13 Gyr is also shown.}
\label{figure1}
\end{figure}

\section{The value of $Y_{\rm GGC}$}

Figure~\ref{figure1} displays the observed $R$ values as a function of
the cluster metallicities, on both the CG97 and ZW84 metallicity scales.
The solid line denotes the theoretical values computed for an age of
13 Gyr and an initial $Y$=0.245. As already discussed in Paper~I,
theoretical models predict an almost constant trend of $R$ with
[Fe/H], with a discontinuity located between [Fe/H]$\sim -$1.15 and 
[Fe/H]$\sim -$0.85. The abrupt increase of $R$ in this metallicity
range is due to the fact that the RGB bump, previously located at
brightnesses larger than the ZAHB, moves below the ZAHB level with
increasing metallicity, thus causing a decrease in the number of RGB
stars brighter than the ZAHB (see, e.g., Salaris et al.~2002). 
As explained in Paper~I, it is only in
this metallicity range that the theoretical calibration of $R$ as a
function of $Y$ does depend on the assumed age, due to the fact
that the RGB bump brightness is also a function of age. The net effect is
to decrease the theoretical value of $R$ at a given $Y$ when the age
decreases. 

Outside this narrow metallicity range a variation of the
age between, i.e. 14 Gyr and 8 Gyr, produces a change in the inferred
value of $Y$ by less than 0.001. This is why in this case we can
safely neglect the precise individual values of the cluster ages 
in this metallicity range, and just assume a common reasonable age. 
Just to show the consistency of our GGC isochrone ages with
independent estimates of the age of the universe (e.g. the WMAP
results), we have determined the age of three clusters spanning
approximately the entire [Fe/H] range of the GGC system, i.e., 
47~Tuc, M~3 and M~15; if we consider, e.g., the ZW84 metallicity
scale, these clusters span the range between [Fe/H]=$-$2.10 and $-$0.7. 
In more detail, we have compared the accurate empirical estimate of
the $\Delta V$
parameter values (e.g. the $V$ magnitude difference between the Turn Off and
ZAHB level, which is a strong function of the cluster age and weakly affected
by metallicity) provided by Rosenberg et al.~(1999), 
with the corresponding values obtained from our isochrones (transformed
to the $V$ Johnson photometric system using bolometric corrections to
the $V$ band homogeneous with the colour transformations 
discussed before). We obtained ages of 11.9$\pm$1.2 Gyr, 11.5$\pm$0.6
Gyr, 11.3$\pm$1.1 Gyr for,
respectively, 47~Tuc, M~3 and M~15. These ages are consistent
with a common value between 11 and 12 Gyr, and consistent also with 
an age of the universe of 13.7 Gyr as estimated from the WMAP results.
Using the CG97 metallicity scale in place of the ZW84 one has only a 
small effect on these ages, of less than 1 Gyr.
We reiterate again that the exact value of the age assumed for the
clusters in our analysis does not influence at all our 
results about $Y_{GGC}$.

\subsection{Estimate of $Y_{\rm GGC}$ using the ZW84 metallicity scale}

\begin{figure}
     \resizebox{\hsize}{!}{\includegraphics{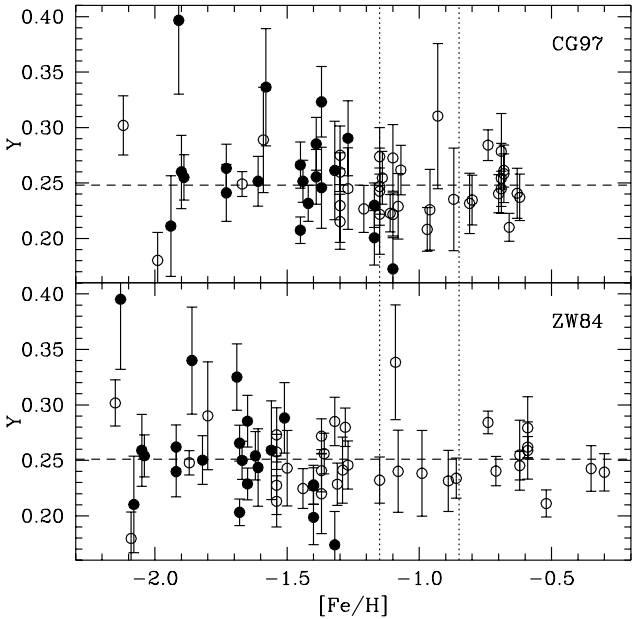}}
      \caption{He abundance values as a function of [Fe/H] for the 57
      clusters analyzed. The two panels refer to the two metallicity scales
      adopted in our analysis. The He abundances determined for the
      clusters with [Fe/H] in the range comprised between the two vertical dotted
      lines are affected by the choice of the cluster age. The 
      horizontal dashed line denotes
      the most probable value for $Y_{\rm GGC}$ (see text for
      details). Filled circles denote clusters with $HB_{type}\geq$0.8.
              }
         \label{figure3}
   \end{figure}

We have first considered the ZW84 metallicity scale. The individual He
abundances with associated 1$\sigma$ errors are displayed in the lower
panel of Fig.~\ref{figure3}. There are 6 clusters in the 
metallicity range affected by the assumed value of the cluster age; 
GGCs in this
metallicity range do show a large age spread (e.g., Rosenberg et
al.~1999; Salaris \& Weiss~2002), towards values lower than the 
common age of the more metal poor objects. This biases the 
initial He abundance inferred with our assumption of a constant 
age (13 Gyr) towards lower values, unless the individual cluster age 
is precisely known and accounted for. These
6 clusters have been therefore excluded from our analysis; when considering  
the whole remaining sample of 51 clusters, we did not
find any statistically significant relationship between $Y$ and
[Fe/H]. We obtain $\delta Y/\delta$ [Fe/H]=$-0.01\pm 0.01$, a ratio 
different from zero by less than its associated 2$\sigma$ error. 

Because there is no dependence of $Y$ on [Fe/H], we may
assume that any spread in the He abundance among the clusters is not
due to Galactic chemical evolution. To investigate 
the existence of an intrinsic abundance spread uncorrelated with
[Fe/H] and possible asymmetries (which can skew the best estimate of the
initial GGC He abundance when using straight averages or weighted averages)
of the He abundance distribution around the modal value,
we have performed the following analysis.
Instead of using histograms with the cluster He abundance
distribution, we determined a continuous probability distribution
function (PDF), following a procedure employed by
Sakai et al.~(1996) for the luminosity function of RGB stars.
More in detail, the He abundance probability density $\Psi(Y)$ 
for a generic value of $Y$ is determined by 
replacing the discretely distributed individual abundances by their
corresponding Gaussian functions (in the hypothesis of Gaussian errors
on the individual abundances), according to the expression

\begin{equation} 
{\Psi(Y)}= \sum_i \ \frac{1}{\sigma_{i}\sqrt{2\pi}}\exp 
          \left(-\frac{(Y_i-Y)^{2}}{2\sigma_{i}^{2}}\right)$$
\label{eq1}
\end{equation} 

\noindent
where $Y_i$ and $\sigma_{i}$ are the He abundance and associated error
of the $i$th cluster. 
$\Psi(Y)$ corresponds to a sum of normalized Gaussian functions, such
that a star with smaller error is represented by a Gaussian that is more
peaked. If $Y$ is the same for all clusters, the PDF
should correspond to a Gaussian function with dispersion $\sigma$
due only to the individual measurement errors.

\begin{figure}
     \resizebox{\hsize}{!}{\includegraphics{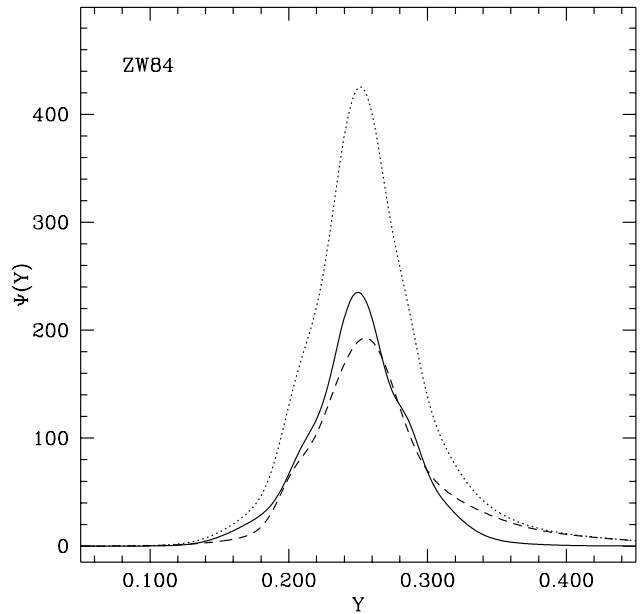}}
      \caption{Probability Distribution Function (PDF) for 
the He abundance in our GGC
sample when using the ZW84 metallicity scale. The dotted line
represents the PDF for the 51 clusters with [Fe/H] in the range not
affected by the cluster age (see text); the solid line represents the
subsample with $HB_{type}<$0.8, while the dashed line represents 
the subsample with $HB_{type}\geq$0.8. The individual PDFs are not
normalized, but for each value of $Y$ the sum of
the PDFs for the two subsamples selected on the basis of
their HB colour is equal to the PDF for the whole sample.
              }
         \label{figure4}
   \end{figure}

Figure~\ref{figure4} displays the PDF for the whole sample, with an
arbitrary normalization. The
peak (i.e. the most probable value of the He abundance for the entire
sample) corresponds to $Y$=0.251, and the 1$\sigma$ dispersion of the
PDF is equal to 0.038; these same values are found when
considering the distribution of the individual cluster abundances.

To check if this observed dispersion is compatible with the cluster individual
errors we have performed, as in Paper~I, the following test.
For each individual cluster we have calculated      
a set of synthetic He abundances by randomly generating -- using  
a Monte Carlo procedure -- 10000 abundance values,  
according to a Gaussian distribution with mean value equal to the observed  
most probable value, and $\sigma$ equal to the     
individual He abundance errors. This is repeated for all     
clusters in the selected sample and the 10000 values for each     
individual clusters are combined to produce an ``expected''     
distribution for the entire cluster sample,  
on the assumption that the detected He abundance dispersion is not  
intrinsic, but due just to the individual errors, assumed Gaussian.        
The F-test was then applied to determine if this ``expected'' distribution, which has
an approximately infinite number of elements, 
displays a variance that is statistically consistent with the observed 
distribution of 51 objects. 
We state that a $Y_{\rm GGC}$ range does exist     
if the probability that the two distributions have different variance     
is larger than 95\%. 

When applying the F-test to our data, we find that this probability is above 
99\%, therefore we can formally conclude that the observed dispersion
is incompatible with the individual error bars. Also the shape of the
distribution is different from Gaussian, as shown by the value of the
kurtosis, which is equal to 2.8.

To investigate this matter further, 
we made a cluster selection on the basis of
their HB colour. We considered the value of the ratio
$HB_{type}$=(B-V)/(B+V+R) (Lee et al.~1994), 
where B, V and R denote the number of HB
stars, respectively, 
bluer than the RR Lyrae instability strip, inside the strip and
redder than the instability strip. The values of $HB_{type}$ are from
the catalogue by Harris~(1996), with the exception of NGC~5634, NGC~5946,
NGC~6273, NGC~6284, NGC~6388, NGC~6441, NGC~6569, NGC~6642, which are
not present in Harris' catalogue. For these clusters we considered the
instability strip boundaries of 
Bono et al.~(1995; 1997), and determined the
number of stars at the blue and red side of the strip (appropriately
reddened according to the individual reddenings given in Harris' catalogue). 
We then computed the value of $HB_{type}$ by considering 
V equal to zero, since we do not have an
estimate of the number of RR Lyrae stars for these clusters. In this way we
have only an approximate upper limit for the value of $HB_{type}$, 
which, as we will see in the following, is enough for our purposes.
By considering various sample selections according to 
the value of $HB_{type}$, we have determined that for the 30 objects with 
$HB_{type}<$0.8 the F-test gives a probability lower than 95\%
that there is an intrinsic spread in the initial He abundances.
The observed 1$\sigma$ spread -- equal to 0.027 --  is therefore compatible
with the individual errors. 
Clusters with $HB_{type}$ in this range 
(see Fig.~\ref{figure3}) still cover the entire [Fe/H] interval spanned by
the whole sample; again, they do not show any statistically
significant trend of He abundance with [Fe/H] 
($\delta Y/\delta$ [Fe/H] =0.002$\pm 0.010$). 
The He abundance of
this sample is therefore formally constant and can be considered to be
an estimate of $Y_{\rm GGC}$. The PDF 
for this sample -- displayed in Fig.~\ref{figure4} -- 
provides a most probable value $Y$=0.250.
It is important to notice that the PDF is very symmetric, which 
is confirmed by the fact that
the weighted average of the individual abundances is equal to 
0.251$\pm$0.003, fully consistent with the most probable value
provided by the PDF. The kurtosis is equal to only $-$0.09, again
in agreement with the fact that the PDF can be well approximated by a
Gaussian function. 
If we include in this sample with $HB_{type}<$0.8 also 
NGC~5634, NGC~6273 and NGC~6284,
for which the computed values of $HB_{type}$=0.91, 0.97 and 0.83 are
only upper limits, the previous results are unchanged.

This result means that the spread 
found for the whole sample, shown to be inconsistent with the
individual measurement errors, must be due to the
21 clusters with $HB_{type}\geq$0.8; in fact, they show 
a 1$\sigma$ abundance
spread of 0.05 around the most probable value of the associated PDF 
$Y$=0.255 (see Fig.~\ref{figure4}). 
This abundance spread is much larger than for 
clusters with redder HB colours, in spite of the fact that there is no
trend in the error on the individual He abundances with
respect to $HB_{type}$. The F-test statistics confirms that the
abundance spread is inconsistent with the individual measurement errors
at the level of more than 99.999\%.

If we do not include NGC~5634, NGC~6273 and NGC~6284 in the cluster
sample with  $HB_{type}\geq$0.8, 
the intrinsic spread is confirmed with the same confidence level.

For the 21 clusters with $HB_{type}\geq$0.8 we do not find a
correlation between their He abundance and $HB_{type}$; 
surprisingly, we obtain a correlation with [Fe/H], with a slope
significant at more than 2$\sigma$ level, i.e., 
$\delta Y$/$\delta$[Fe/H]=$-0.10\pm$0.04. This correlation is
unexpected because it does not exist for clusters with redder HB
types. We will see that it is also absent from the subsample with blue
HBs if we use the CG97 [Fe/H] scale, and we believe it is just an
artifact due to the uncertainty on the GGC metallicity scale.

  \begin{figure}
     \resizebox{\hsize}{!}{\includegraphics{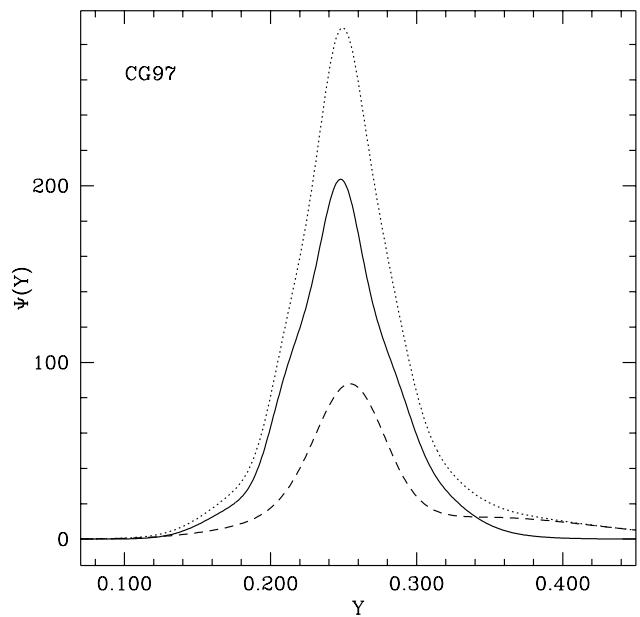}}
      \caption{As in Fig.~\ref{figure4}, but for the CG97 [Fe/H] scale
               (see text).
              }
         \label{figure5}
   \end{figure}

It is therefore confirmed that the 
large abundance spread incompatible with the measurement errors
found for the whole
sample is due to these clusters with the bluest HB colours. They are
the most affected by the possible presence of stellar subpopulations 
with enhanced He, as well as by the dependence of the HB evolutionary
timescale on the stellar mass. Both effects tend to 
increase the He abundance estimated with our calibration. 
For these blue HB clusters we find that the peak of the 
asymmetric PDF is located at an abundance larger
than the value obtained for clusters with redder HB, which display 
an abundance spread compatible with the measurement errors.
Figure~\ref{figure4} also shows how the high abundance tail of
the PDF has more weight than for clusters with redder HB
(the value of the kurtosis is equal to 1.0 for this subsample of blue
HB GGCs). We therefore interpret the He abundance spread 
being incompatible with the measurement errors
found in this group of blue HB clusters as the effect of the increasing
evolutionary timescale for very blue HB objects, eventually coupled to
the presence of He enriched stellar subpopulations. This conclusion is
reinforced by the results of the next section.

To avoid these problems with blue HB clusters 
we will therefore consider -- as said before -- the sample of 21 clusters 
with $HB_{type}<$0.8 for the best estimate of $Y_{GGC}$.
Since the observed spread is compatible with the measurement errors,
one possibility -- which is what we have implicitly assumed in Paper~I -- 
is to consider that the intrinsic dispersion of the individual $Y$ is
zero; this assumption is fully compatible with the F-test analysis.
In this case the best estimate of $Y_{GGC}$ is 
the weighted average (which also agrees with the mode of the
associated PDF, as shown before) that provides $Y_{\rm GGC}$=0.251$\pm$0.003.

Another possibility is that there exists an intrinsic abundance
spread not detected by the F-test, because it is masked by the
individual errors. We can investigate this issue more deeply by
assessing what intrinsic dispersion in the individual He abundances is
compatible with the observed distribution of $Y$. Applying the
F-test to the actual sample of $Y$ abundances and to synthetic samples 
computed accounting for the actual individual errors plus 
increasingly large abundance spreads
(in case of Gaussian and uniform intrinsic spread, 
although of course the mathematical form of the
hypothetical abundance spread is unknown), we obtained the somewhat
expected result that an intrinsic 1$\sigma$ dispersion up to 0.025
is still allowed by present data. This means that 0.025 is an upper
limit to the possible intrinsic spread within our 21 object sample.

As a further test we selected a subsample of 12 clusters with
individual errors lower than 0.025. These clusters span the whole 
metallicity range covered by the full sample, and their individual He
abundances do not show any trend with metallicity.
The corresponding abundance distribution 
shows the same mode and weighted mean as the 21 object sample, 
and a reduced 1$\sigma$ dispersion of 0.021, again fully compatible with the
individual errors (as assessed through an F-test). 
Following the same procedure outlined above, we
found that the upper limit to the intrinsic 1$\sigma$ spread allowed by this
sample is equal to 0.019.
It should be realized that 
this subsample with smaller individual errors, still containing 
a sizable number of objects
spanning the entire metallicity range, has the same mode and
weighted mean and shows an He abundance distribution 
with a lower dispersion fully compatible with the individual errors.
This allows one to safely reduce the upper limit of a possible
intrinsic 1$\sigma$ spread to 0.019.

From our data we can therefore conclude that, 
if one allows for an intrinsic abundance
spread, the most probable value $Y_{\rm GGC}$
is well represented by the weighted average given above, since it
is practically coincident with the mode of the sample PDF. 
An upper limit of 0.019 can then be associated to 
the 1$\sigma$ dispersion around $Y_{\rm GGC}$ allowed by present data.

\subsection{Estimate of $Y_{\rm GGC}$ using the CG97 metallicity scale}

We have repeated our abundance analysis considering 
the CG97 metallicity scale  (see Fig.~\ref{figure3}). 
In this case, there are 15 clusters in the metallicity range affected
by the cluster age, about 25\% of the whole sample. 
In the following we will consider only the 42 clusters 
with [Fe/H]$> -$0.85 and [Fe/H]$< -$1.15. For these clusters there is
again no statistically significant trend of He abundance with [Fe/H]
($\delta Y/\delta$ [Fe/H]=$-0.017\pm 0.014$).

The PDF for the He abundance of this sample is displayed in 
Fig.~\ref{figure5}. The most probable value for the He abundance is 	
$Y$=0.250 and the 1$\sigma$ dispersion is equal to 0.04. 
When applying the F-test to this sample, we obtain a statistically
significant He abundance spread at more than 99\% confidence level. 
The shape of the distribution is also very different from a Gaussian
one, with a value for the kurtosis equal to 3.1.
As for the ZW84 metallicity scale, we find that 
the 22 clusters with 
$HB_{type}<$0.8 do not show a statistically significant 
He abundance spread, nor a trend of He abundance with [Fe/H] 
($\delta Y/\delta$ [Fe/H]=$-0.002\pm 0.014$); for
these 22 clusters we obtain a most probable value
$Y$=0.248. The PDF is again highly symmetric and the value of the 
weighted average
of the individual abundances, $Y$=0.246$\pm$0.004, fully agrees with 
the location of the peak of the PDF; the kurtosis is negligible, 
equal to $-$0.03.

For the 20 clusters with $HB_{type}\geq$0.8, the abundance spread
is statistically significant (at more than 99.999\% confidence level), the most
probable value for the He abundance being equal to $Y$=0.251, and there is
no trend of the individual abundances with either [Fe/H] or 
$HB_{type}$. As for the ZW84 metallicities, the
bluest HB clusters have He abundances shifted to higher values, and
again the high He abundance tail of the PDF has substantially more
weight than for the case of the clusters with redder HB (the value of the
kurtosis is equal to 1.5). 
The lack of correlation of the individual abundances with [Fe/H] for
this subsample strongly suggests that the correlation found with ZW84 
metallicities is spurious, due to the actual uncertainty on the GGC
[Fe/H] scale.

As for the ZW84 metallicities, we can consider the
subsample of 22 clusters with $HB_{type}<$0.8 to provide the best
estimate of $Y_{GGC}$. As before, the F-test analysis indicates that 
either there is no intrinsic spread in the individual He abundances,
or some intrinsic spread is masked by the individual measurement
errors. In the first case one can use the weighted average 
$Y_{\rm GGC}$=0.246$\pm$0.004 as the best estimate of $Y_{GGC}$. In
the second case, following exactly the same steps as for 
the ZW84 metallicities, we find that the weighted average 
(again consistent with the mode of the PDF) also provides 
the most probable value for $Y_{GGC}$,  
with a 0.019 upper limit to the 1$\sigma$ dispersion
allowed by the present data.

\section{Summary and conclusions}

In this paper we have presented a detailed study of the initial He abundance
of the GGC system. We have used the most extensive and homogeneous
database of $R$-parameter values to date, and recent updated stellar
evolution models. As in Paper~I, we do not find any statistically
significant correlation between the individual cluster He abundances 
and [Fe/H]. 
This suggests a very homogeneous value of $Y$ for the 
GGCs, with practically no He abundance evolution over the entire
[Fe/H] spanned by the GGC system, at variance with results on the
evolution of He in extragalactic HII regions, which measure a 
gradient of He with Oxygen (hence with [Fe/H]). 
This result on the evolution of He in GGCs strengthens similar conclusions
reached in Paper~I, and it has to be taken into account in studies of
Galaxy formation mechanisms. Note that 
Sandquist~(2000) also found a constant $Y$ for the GGC system 
using two independent indicators of relative He abundances, 
namely the $\Delta$ parameter
(difference in magnitude between HB and Main Sequence) and the RR
Lyrae mass-luminosity exponent $A$.

As for the existence of an intrinsic spread of the cluster initial
$Y$, we found that objects with $HB_{type}<$0.8 -- irrespective of
their [Fe/H] -- do show a remarkably homogeneous and symmetric
abundance distribution, 
whose spread is fully compatible 
with the measurement errors. 
One can therefore conclude that either the He abundance is the
same for all clusters, without any intrinsic dispersion, or that an
eventual dispersion is masked by the individual errors. In this
case we obtain that our data allow a 0.019 upper limit to a possible 
1$\sigma$ intrinsic dispersion.
On the other hand, clusters with blue HB colours 
($HB_{type}\geq$0.8) show a spread in their initial $Y$
incompatible with the individual error bars, skewed towards
values higher than those for clusters with redder HB. 
We interpret this result as due to the increase of evolutionary timescales along the HB
phase for progressively bluer HB stars, which causes an overestimate of $Y$ when
using our calibration. This effect may be coupled to the possible presence of a
subpopulation of He enhanced stars in at least some GGCs (see D'Antona et al.~2002),
which also leads to an overestimate of $Y$ (at most by 0.01 -- 0.02)
when determined from our calibration.

As already noticed, for clusters with $HB_{type}<$0.8 
we obtain a distribution of
$Y$ values very close to Gaussian and a 1$\sigma$ dispersion consistent
with the measurement errors. 
This occurrence goes against the possible
existence of a sizable spread towards higher $Y$ values due to subpopulations
of He enriched stars, as in blue HB clusters; 
in fact, in case redder HB clusters, the effect of
these subpopulations is at most at the level of $\sim$0.005, and
probably smaller. It is also reasonable to assume that these subpopulations
might not be present in all clusters, and the net effect on our 
estimates of $Y_{\rm GGC}$ 
should therefore be negligible. 
An extreme 
possibility is that all clusters with $HB_{type}<$0.8 are affected in  
the same way by this enhancement, whose maximum effect would be to bias
our $Y_{\rm GGC}$ estimate by $\sim$0.005 towards too high values. 

We conclude by providing a best estimate  
for $Y_{\rm GGC}$, including the effect of the still uncertain [Fe/H] 
scale and the sources of systematic errors discussed in Paper~I (i.e., 
the error on the $^{12}$C$(\alpha,\gamma)^{16}$O 
reaction rate and the method adopted to suppress the breathing pulses
during the final stages of central He burning).
We have considered as reference value the  
$Y_{\rm GGC}$ abundance determined with the ZW84 [Fe/H] scale. 
Starting with this reference $Y_{\rm GGC}$  
we have generated a set of 100000 synthetic  
He-abundance values, by applying (through a Monte Carlo simulation)  
to the reference value a set of random and systematic errors, 
according to a given probability distribution.  
In particular, random errors have been modeled according to a Gaussian  
distribution with mean value equal to the reference one, and 1$\sigma$  
dispersion equal to the corresponding random error on $Y_{\rm GGC}$.  
The systematic uncertainties due to the choice of the [Fe/H] scale  
(which causes a decrease of $Y_{\rm GGC}$ by 0.005 with respect to the  
reference value), $^{12}$C$(\alpha,\gamma)^{16}$O reaction rate  
(variation by $\pm$0.008), and breathing pulses suppression technique  
(increase by 0.003) have been modeled using a uniform  
distribution spanning the appropriate range.  
  
The mean value for the final synthetic distribution of He abundances  
is $Y_{\rm GGC}$=0.250$\pm$0.006. An intrinsic dispersion
with a firm 1$\sigma$ upper limit of 0.019 around this value of
$Y_{\rm GGC}$ is a priori possible given the observational errors.
This estimate of $Y_{\rm GGC}$ 
is in good agreement with the primordial He abundance 
inferred from the CMB in conjunction with Big-Bang 
nucleosynthesis computations. Within the respective
1$\sigma$ errors this value is also in agreement 
with the results from Paper~I.

\begin{acknowledgements}
We warmly thank R. Gratton, G. Steigman and A. Irwin for interesting 
discussions
on this topic and useful suggestions, and an anonymous referee for
very pertinent remarks.
M.S. wishes to thank past and present Rockets for the inspiration they
provided during all these years, and Toby Moore for useful discussions. 
This work has been partially supported by the Italian Ministero
dell'Istruzione e della ricerca (PRIN2002 and PRIN2003) and by
the Agenzia Spaziale Italiana.

\end{acknowledgements}


\begin{thebibliography}{}

\bibitem[]{} Alexander, D.R., \& Ferguson, J.W. 1994, ApJ, 437, 879

\bibitem[]{} Angulo, C., et al. 1999, Nucl.Phys.A, 656, 3  

\bibitem[]{} Bessel, M.S., Castelli, F., \& Plez, B. 1998, A\&A, 333, 231

\bibitem[]{} Bono, G., Caputo, F., \& Marconi, M. 1995, AJ, 110, 2365

\bibitem[]{} Bono, G., Caputo, F., Cassisi, S., Incerpi, R., \&
Marconi, M. 1997, ApJ, 483, 811

\bibitem[]{} Burles, S., Nollett, K.M., \& Turner, M.S. 2001, ApJ,     
552, L1  

\bibitem[]{} Buzzoni, A., Fusi Pecci, F., Buonanno, R., \& Corsi,     
C.E. 1983, A\&A, 128, 94  

\bibitem[]{} Caputo, F., Martinez Roger, C., \& Paez, E. 1987, A\&A,     
183, 228 

\bibitem[]{} Cassisi, S., Castellani, V., degl'Innocenti, S., \&
Weiss, A. 1998, A\&AS, 129, 267

\bibitem[]{} Cassisi, S., Salaris, M., \& Irwin, A.W. 2003, ApJ, 588, 862

\bibitem[]{} D'Antona, F., Caloi, V., Montalban, J., Ventura, P., \&
Gratton, R.G. 2002, A\&A, 395, 69 

\bibitem[]{} Harris, W.E. 1996, AJ, 112, 1487

\bibitem[]{} Iben, I. Jr. 1968, Nature, 220, 143

\bibitem[]{} Iglesias, C.A., \& Rogers, F.J. 1996, ApJ, 464, 943

\bibitem[]{} Izotov, Y.I., \& Thuan, T.X. 1998, ApJ, 497, 227

\bibitem[]{} Izotov, Y.I., \& Thuan, T.X. 2003, ApJ, in press (astro-ph/0310421)

\bibitem[]{} Kunz, R. et al. 2002, ApJ, 567, 643  

\bibitem[]{} Lee, Y-,W., Demarque, P., \& Zinn, R. 1994, ApJ, 423, 248

\bibitem[]{} Michaud, G., Vauclair, G., \& Vauclair, S. 1983, ApJ,     
267, 256 

\bibitem[]{} \"Odman, C.J., Melchiorri, A., Hobson, M.P., \& Lasenby     
A.N. 2002, Phis.Rev.D, 67, 083511

\bibitem[]{} Origlia, L., \& Leitherer, C. 2000, AJ, 119, 2018

\bibitem[]{} Olive, K.A., Skillman, E.D. \& Steigman, G. 1997, ApJ, 483, 788

\bibitem[]{} Piotto, G., et al. 2002, A\&A, 391, 945

\bibitem[]{} Potekhin, A.Y. 1999, A\&A, 351, 787 

\bibitem[]{} Pryke, C., et al. 2002, ApJ, 568, 46 


\bibitem[]{} Riello, M., et al. 2003, A\&A, 410, 553

\bibitem[]{} Rosenberg, A., Saviane, I., Piotto, G., 
\& Aparicio, A. 1999, AJ, 118, 2306

\bibitem[]{} Sakai, S., Madore, B.F. \& Freedman, W.L. 1996, ApJ, 461, 713 

\bibitem[]{} Salaris, M., Cassisi, S., \& Weiss, A. 2002, PASP, 114, 375

\bibitem[]{} Salaris, M., \& Weiss, A. 1998, A\&A, 335, 943 

\bibitem[]{} Salaris, M., \& Weiss, A. 2002, A\&A, 388, 492

\bibitem[]{} Sandage, A., \& Wildey, R. 1967, ApJ, 150, 469

\bibitem[]{} Sandquist, E. L. 2000, MNRAS, 313, 571 

\bibitem[]{} Sievers, J.L., et al. 2003, ApJ, 591, 599

\bibitem[]{} Spergel et al. 2003, ApJS, 148, 175

\bibitem[]{} Zoccali, M., Cassisi, S., Bono, G., Piotto, G., Rich,        
R.M., Djorgovski, S. G. 2000, ApJ, 538, 289    
   
\end{thebibliography}
\end{document}